\documentclass[doc]{apa6}

\usepackage[utf8]{inputenc}
\usepackage[american]{babel}
\usepackage{float}
\usepackage{csquotes}
\usepackage{array}
\usepackage{amsmath}
\usepackage{tabularx}
\usepackage{spverbatim}
\usepackage[style=apa,sortcites=true,sorting=nyt,backend=biber]{biblatex}
\DeclareLanguageMapping{american}{american-apa}
\title{Know your population and know your model:\\Using model-based regression and post-stratification to generalize findings beyond the observed sample}
\shorttitle{Multilevel Regression and Poststratification in psychological research.}
\twoauthors{Lauren Kennedy*}{Andrew Gelman}
\twoaffiliations{Population Research Center and Department of Statistics, \\Columbia University, New York.}{Department of Statistics and Department of Political Science, \\Columbia University, New York.}
\authornote{We thank the U.S. Office of Naval Research, National Science Foundation, and Institute of Education Sciences for their partial support of this work. We also thank Jessica O'Rielly and Matthijs V{\'a}k{\'a}r for their thoughtful comments on this manuscript.\\ Preprints of this work have been posted on arxiv.org, and ideas around this work have been presented at the Australian Mathematical Psychology meeting (2019) and the 2019 Society of Experimental Social Psychology. \\*Corresponding author. Email: \texttt{lauren.kennedy729@gmail.com}}

\abstract{Psychology research focuses on interactions, and this has deep implications for inference from non-representative samples.  For the goal of estimating average treatment effects, we propose to fit a model allowing treatment to interact with background variables and then average over the distribution of these variables in the population.  This can be seen as an extension of multilevel regression and poststratification (MRP), a method used in political science and other areas of survey research, where researchers wish to generalize from a sparse and possibly non-representative sample to the general population. In this paper, we discuss areas where this method can be used in the psychological sciences. We use our method to estimate the norming distribution for the Big Five Personality Scale using open source data. We argue that large open data sources like this and other collaborative data sources can be combined with MRP to help resolve current challenges of generalizability and replication in psychology.}


\begin{document}

\maketitle

Psychology is all about people, and because people are so wonderfully heterogeneous, psychology has to consider the impact of interactions. Not even randomization can save us from heterogeneity. Even in studies that only claim to investigate whether an effect ``exists,'' the expected heterogeneity of the effect should be considered when interpreting the results. At the same time, some studies are concerned with effects that hold in some broader population. If our sample isn't representative (as many psychological research samples are not), and the effect is heterogeneous, how can we estimate this effect in the population?

This challenge is not new to statisticians. Traditionally, survey weighting has been employed to account for differences between sample and intended population from design and nonresponse. However, it is uncommon for participants in a psychology experiment to be chosen using any formal sampling procedure. Convenience samples (or non-probability samples) dominate the field, which makes it difficult to construct classical sampling-based. Even random samples from surveys are rarely actually random in the real world of nonresponse. This is well known and has led to survey weights being constructed based on adjusting for estimated probability of inclusion in the sample. Without random sampling, this problem grows even more difficult. Often in psychology we rely on convenience samples, such as first year undergraduates, kind community members, or (more recently) Amazon Mechanical Turk workers and other crowd-sourcing alternatives. These convenience samples rarely represent the population that we are interested in, and they can differ in important ways from underlying populations of interest.

Throughout this paper we argue that the statistical technique known as multilevel regression and poststratification (MRP; \cite{little1993post, park2004bayesian,gelman1997poststratification}) could be applied to convenience samples in psychology. This method allows the researcher to infer quantities in the \textit{population} from a sparse and possibly non-representative \textit{sample}, combining two ideas in the survey research literature:  small-area estimation and nonresponse adjustment. MRP is popular and is a continuing subject of research within the political science literature (see, for example, \cite{ghitza2013deep,lax2009should,Si_2017}) and also been introduced in public health \parencite{downes2018multilevel}. \cite{wang2015forecasting} demonstrate the effectivness of MRP for a large non-probability political poll.

In psychology we already know about experimental design. In this paper we are not discussing an alternative to randomization, nor are we considering the challenge of generalizing to new experimental conditions not in the existing study. Instead we focus on generalizing about the treatment at hand as it would apply to a wider population beyond people in the sample. While this is a relatively uncommon adjustment within psychology, political science endeavours such as polling consider differences between sample and the population much more often. One reason for this is that the population of interest (e.g., voters or the general adult population) is clearly defined. The other is that when different polls disagree, we can ultimately compare to events such as elections that provide ground truth.

To perform our extrapolations we make two types of assumptions. First, we make statistical model assumptions in terms of variables included, priors (if any) used, and the type or form of the model. In particular, if we are interested in extrapolations of treatment effects, it is important to include interactions between the treatment and the person-level variables that capture key differences between sample and population.
Second we make assumptions of equivalence - that the people unobserved are the same as the observed once we have adjusted sufficiently. If we adjust on age group and gender, equivalence means that people within a specific age x gender group would have the same expected difference given an intervention (with some random variation). 

MRP and other survey adjustments are not widely used to analyze experimental data in psychology. Randomization of treatment assignment is thought to allow us to estimate the average treatment effect.  However, in the presence of interactions between demographic characteristics and the quantities of interest (which are typically the object of study in psychology experiments), the average treatment effect is uninterpretable without reference to a population, hence adjustment for non-representativeness of the sample again becomes necessary. Even in experiments that are only concerned with whether an effect exists \parencite[the ``what can'' argument of][]{mook1983defense}, heterogeneity can explain when a study doesn't find an effect (even when there is one for some groups), or why an effect didn't replicate after being observed once, a point highlighted by \textcite{henry2008college}.

In explaining how MRP has potential to be useful for generalizing research findings in psychology, we first discuss in high-level language what we mean by multilevel modeling and poststratification, and how the two methods combine to be such a useful tool. Then we describe some caveats with MRP, before using an open data set measuring scores on the Big Five Personality Scale to demonstrate an application of MRP. We conclude with a discussion of the limitations and active research currently conducted in this area.

\section{What is MRP?}

Multilevel regression and poststratification combines two statistical techniques to (a) quantify the relationship between some outcome variable of interest and a number of predictors, and (b) obtain generalizable inferences by adjusting for known discrepancies between sample and population. Similar approaches use alternative models with poststratification \parencite[e.g.,][]{caughey2019public, gao2019improving}. The important point is that the model uses some sort of regularization or partial pooling to obtain stable estimates from relatively small samples. Here we focus on what might be the most common among psychologists, mixed effects models, to explain how this regularization works. 

A mixed effects model is similar to a traditional regression (where some outcome variable $y$ is modeled as a function of a set of predictors $x_1,x_2,x_3,\dots,x_m$), but a mixed effects model breaks these predictors into two sets; constant and varying effects. We avoid the terms `fixed' and `random' here because they are given different meanings in different contexts; see \cite{gelman2005analysis}.

In the case of MRP, the technique advocates for using varying effects for predictors such as education, race/ethnicity, state, and age category that take on multiple levels in the data.  We do not restrict them to be used for multiple observations per individual (as in the traditional use of multilevel models in psychology). We demonstrate how multilevel modeling differs from classical least-squares regression or ANOVA with a simple hypothetical example. For a more detailed description, we recommend \textcite{sorensen2015bayesian}. We will also build on the notation of \textcite{gelmanhill}, which is commonly used in the MRP literature. For reader ease, we begin with a hypothetical example where multilevel models have often been used. Say you have test scores from a sample of students each belonging to one of $K$ schools, and you are interested in predicting scores $y$ from school $k$. How is multilevel regression, with varying intercepts for school, different from least-squares regression including school indicators?

The classic model setup for including school effects would be to create $K$ binary variables, denoted $d_k$. (An alternative parameterization is to create $K - 1$ indicator variables with the $K$th replaced by the intercept. We formulate the model with $K$ predictors as it allows easier formulation as a varying effect.)   Each variable indicates whether the student belongs to school $k$. We could then fit the following non-multilevel model:
\begin{equation}
    \label{eq:constant}
    \begin{gathered}
    y = \beta_1*d_1 + \beta_2*d_2 + \beta_3*d_3 + \dots + \beta_{K}*d_{K} +\epsilon, \\
    \epsilon \sim N(0,\sigma_y^2)
    \end{gathered}
\end{equation}
If a student is in school $7$, for example, then $d_7 = 1$ and all other $d_{k; k\neq{7}} = 0$. This means that the above equation would simplify to:

\begin{equation}
    y = \beta_7 + \epsilon,
\end{equation}
for which the estimate would simply be the mean of school $7$. 

In multilevel regression, we would model the intercept for the schools as $\beta_k, k=1,\dots,K$, and then apply a probabilistic or `soft' constraint to the set of $\beta_k$'s such that they are distributed with mean $\mu$ and variance $\sigma$. 

\begin{equation}
    \label{eq:varying}
    y = \beta_k +\epsilon.
\end{equation}
    
\begin{equation}    
    \beta_k \sim normal(\mu_k, \sigma_\beta).
\end{equation} 

The two models are similar in that each school is modeled as having a different mean level of scholastic ability. The difference is the amount of information that is shared between the levels. In the first formulation, the test scores in each school are modeled independently of other schools. In the second formulation, the test score component for each school uses information observed test scores at other schools. With multilevel modeling, the amount of shared information forms a continuum, ranging from no pooling (Equation~(\ref{eq:constant})) to full pooling, which would correspond to a model with an intercept that is the same for each school (equation below). \textcite{gelman2006bayesian} describe this continuum more formally.

\begin{equation}
    \label{eq:contant}
    y = \beta_{int} + \epsilon.
\end{equation} 

Multilevel modeling allows us to fit the amount of pooling (through the size of $\sigma_\beta$) with the other parameters in the model. The amount of pooling is also akin to the amount of regularization. More pooling indicates more regularization, less pooling indicates less regularization. Moreover, it provides an avenue to make predictions about new populations or samples. One example of this is in \textcite{weber2018bayesian}. To do this we might need to use a strong prior about the relationship of the observed sample to the sample or population that we would like to generalize to. 

This leads us to the second component of MRP, poststratification. For the school example, we would need a poststratification table that contains the total number of students in each school. We would use the formula obtained from the regression analysis to predict the test scores for each school. To obtain an estimate for the total population of students, we would multiply each school estimate by the number of students in that school, add these all up, and then divide by the total number of students in the population. Mathematically if the school estimate for the $k^{th}$ school is referred to as $\theta_k$, this would be expressed as:

$$\theta_{\texttt{POP}} = \frac{\sum_{k\in K}N_k \theta_k}{\sum_{k \in K}N_k }$$

The steps to MRP are as follows:

\begin{enumerate}
    \item Measure key demographic features in sample during survey collection.
    \item Identify the poststratification table: estimated population counts for each possible combination of these demographic features (each combination is a cell in the table). 
    \item Measure some key quantity in the sample. This is what you would like to estimate in the population. 
    \item To estimate this quantity of interest in the population, use multilevel modeling to predict this quantity using the observed demographic features in the sample.
    \item Estimate the outcome variable each cell of the poststratification table.
    \item Aggregate over cells of the poststratification cells (using the cell size) to obtain population level estimates. 
\end{enumerate}

\subsection{Conditions of data necessary for generalization}
Not all situations are suitable for generalization through MRP. The method is designed to be used in examples where we expect heterogeneity, the heterogeneity is expected to interact with the outcome or manipulation, and the sample is not representative of the population. If all three conditions are met, then there is potential for this approach to be beneficial. 
Even when we do have an example that meets these conditions, the data that we have collected might not be sufficient to model it in this way. First, to model heterogeneity, the data actually need to contain heterogeneity. If the sample is an undergraduate population and the effect is expected to differ between young adults and the elderly, then this method will not be appropriate to estimate a population effect but might be appropriate to estimate an undergraduate effect. If the sample is an undergraduate population and we don’t expect there to be heterogeneity across age or previous research demonstrates there is not, then perhaps the sample is suitable. This is discussed partially in \textcite{smith2018small} in relation to small-N vision studies. 

Although this might seem counterintuitive, in practice no sample, even one obtained by random sampling, can be truly representative on all possible covariates. Although MRP traditionally follows the survey weighting literature to adjust for individual demographics and sampling design; different sets of adjustment variables might be more appropriate when generalizing in psychology. This raises challenges because these adjustment variables may not be known in the population, by applying MRP to fields like psychology will also give an opportunity for the field of survey research to improve the methods currently in place. 

Another necessary condition for MRP is sufficient data to fit these hierarchical models. It’s difficult to give broad recommendations for the required sample size, but we doubt that much would be gained if this method were used with a between-person design with a small sample such as from 50 individuals. When using MRP were to be used with this sample, we expect that the estimates for new groups would be very noisy, or else inferences would depend strongly on prior.

In addition, we recognize that not all research in psychology is about estimating population level effects. In some fields the research question is simply whether an effect is observed or not. In this case MRP is not directly relevant.  That said, if effects are heterogeneous, then not observing an effect in a given study does not mean that this effect doesn’t exist in the wider population. Similarly, observing a “significant” effect in a particular sample could be difficult to replicate in different samples.

\section{What should we adjust for?}

In political science the variables that we expect to adjust for are fairly consistent. Basic demographics such as age, sex, race/ethnicity, education, along with geographic factors such as state and urban/rural/suburban. But what adjustment variables should we use in psychology applications? Basic demographics should be a good starting point, but applying this method in any particular example would benefit from understanding where heterogeniety is expected in terms of the impact of a given intervention. 

This knowledge is typically ingrained in the expertise of the researchers conducting a study. For example, \textcite{sears1986college} needed to have a good understanding of the college student population and the social phenomena he was studying to discuss the implications of a college student sample on the phenomena. However, recently \textcite{simons2017constraints} advocate for the inclusion of a constraints of generality (COG) statement in all empirical work. We believe that such a statement will help the researcher communicate knowledge about expected heterogeneity of effect and provide clues of what to adjust for. Indeed, in their paper they argued that such a statement will make clear where findings are expected to replicate, and encourage other researchers to explore outside of the proposed boundary conditions.

We believe that the constraints on generality statement provides necessary information to move beyond a general statement about replicability to a statistical approach leading to quantitative conclusions (and, as appropriate, large uncertainties) about particular replications or generalizations of interest.  Additionally, the COG statement has been subjected to peer review. Although peer review is not infallible, it does provide some suggestion the COG statement reflects the knowledge and experience of researchers in the field. We discuss the importance of this later in this article.  

\section{Incentives to use MRP}

Having discussed how MRP is formalized, in this section we discuss why such a technique can be so useful. The advantages of MRP align with the contents of an effective COG statement. That is, MRP is useful when we as researchers have formalized a population of interest, have identified key variables that are believed or theorized to impact the outcome variable of interest, and have distinguished differences between the sample and the population using these key variables. By our interpretation of \textcite{simons2017constraints}'s paper, the COG provides a structure to do exactly that. Unsurprisingly given the close relationship between the two, the incentives of MRP mirror those of the COG. 

The COG statement gives the researcher an avenue to consider the population of interest and to consider whether findings from the sample should generalize to the population as a whole. MRP uses population level information to \textit{directly} estimate the variable of interest in said population, including an estimate of uncertainty. Without the COG, which involves the researcher considering what population they hope to generalize to, and the differences between the sample and population that might impact generalizations from the sample to the population, we would not know which variables to include in our multilevel model, nor would we be able to define the population well enough to generalize to it. The COG statement provides this information, so MRP can build upon it to infer things about the population. 

Likewise, while the COG statement provides some basis for the researcher to guess how likely it is that the findings will replicate based on differences between previous sample populations and the current sample, MRP provides a way to estimate the variable of interest in a new sample directly. While dissonance between the MRP estimate and the observed value in a new sample doesn't immediately signal a failure to replicate, it does provide a tool for further research to explore whether there are additional differences between the two samples that might cause failure to replicate. 

Lastly while the COG statement uses researcher intuition and domain-specific knowledge of the field, the multilevel component part of MRP provides an avenue to test and quantify these beliefs. While it might be intuitive that a specific demographic variable might be related, multilevel regression helps to quantify the size of the relationship, leading to better population level predictions. 

By conducting MRP and finding heterogeneity between demographic subgroups, we can also find inspiration for future research. For example, say we wish to estimate mathematical reasoning in the population and find that in our sample gender is a good predictor. The current research project might poststratify using gender to obtain population level estimates of mathematical reasoning, while future research might focus on exploring this relationship in more depth. 

All of these incentives require the researcher to be able to formally state and describe the population that they wish to generalize to, which to us is one of the biggest benefits of the COG statement. In the next section we consider how the COG and MRP might work together in practice. 

\section{Example 1: Tutorial with real data}

In this section we use an open source dataset to demonstrate the mechanics of an MRP approach. Say you are developing some new scale (such as a personality scale) for use in the general public. After validating that it measures what it is intended to, your next step would be to estimate the distribution of this scale in the general population so that an individual's score can be meaningful relative to the greater population. There are a couple of ways to achieve this.

To see how this works in practice, we apply this technique to a large database of responses to the Big Five Personality Scale \parencite{goldberg1992development} collected through the \cite{OCEANdata}. This scale measures five facets of personality; Openness, Conscientiousness, Extraversion, Agreeableness, and Neuroticism. Each subscale is measured with 10 items measured with a five-point likert scale with items scored from 1 to 5 (with some items scored in reverse) so that total scores on each subscale range from 10 to 50. This tutorial is accompanied with a Rmarkdown document (for Openness subscale) and .R file (for all subscales), found in the supplementary materials. 

This dataset contains a convenience sample of $19\,719$ individuals who completed the scale online. Following the scale, participants also were asked to provide basic demographic information, with location information derived through technical information, which we used to subset to US participants specifically. A total of $8\,665$ US participants provided all of the requested information. 

One way to interpret the Big Five is to compare an individual's score in relation to the wider population distribution. To do this, we need a distribution of scores on a representative sample. This is particularly important as there are substantial individual differences in  personality scores, for example across age \parencite{donnellan2008age} and gender \parencite{weisberg2011gender}. 

A convenience sample is rarely representative. In this particular case, our sample was much less like to be male ($M= 34\%$ or $2\,939$) when compared to the wider US population as estimated from the 2012 ACS ($M = 49\%$). The proportion of the sample aged 13--25 is 60\% in our sample but only 20\% in the ACS. Although we do not know the who decided to participate in this study, we can guess from the other demographics (predominantly young women) that at least some portion were undergraduate psychology students. Ideally we would also be able to adjust for education level of our sample but this covariate is not available in our dataset. This is a major limitation of this analysis as it means we are assuming that either our sample does not differ from the population on education level, or that education level is not related to the Big Five. We are also limited in that the survey we are using does not measure race using the same categories as those in the American census. Ideally we would create a mapping between the two measurements, but for the purposes of this tutorial we can still adjust the sample distribution of each of the facets of the Big Five on gender and age group. The accompanying Rmarkdown document demonstrates this analysis for the Openness subscale, which is repeated for the four other subscales in the included R script. 

\subsection{Step 1: Model the outcome in terms of the adjustment variables}

After downloading the data and reverse coding the necessary items, we sum each of the 10 items to get a total score on each subscale of the Big Five. These subscales are the outcomes that we would like to estimate in the population. To do this we need to fit a multilevel model with age and sex as the adjustment variables (predictors). The dataset measures gender (male, female, or other), while the ACS measures sex (male or female). For simplicity, we remove all cases where gender is not stated as male or female. We hope future research will work on more appropriate ways to poststratify gender to the census. Age is broken into six uneven categories; under 18 $(N=1\,903)$, 18--24 $(N=3\,285)$, 25--34 $(N=1\,507)$, 35--44 $(N=847)$, 45--65 $(N=890)$, and 65+ $(N=233)$.  The dataset also measures race/ethnicity, but does not use categories that map easily to those used in the US census so we do not adjust by race/ethnicity.

For each outcome (O, C, E, A, and N) we fit a model in R using brms \parencite{brms}, a package that allows the user to fit fully Bayesian models using standard R formula notation and with enough flexibilty that our model can account for truncation of the outcome variable between 10 and 50. It is possible to perform multilevel modeling without being fully Bayesian, but we find that a Bayesian approach is very natural, especially for accounting for different sources of inferential uncertainty when making predictions. 

For the purpose of readability, we describe the process using one outcome variable---scores on the Openness subscale---but adjust all five subscales in the accompanying code. With the following code we fit a regression model with upper and lower bounds (\texttt{ub=50} and \texttt{lb=10}) with \texttt{O} as the outcome variable, gender as an indicator for female, and \texttt{age\_group} as a varying effect. We specify the data as \texttt{data\_us}. The remaining input specifies computational details, namely that $4$ chains of MCMC will be run, that there are $4$ available cores, and the step size (\texttt{adapt\_delta}) and treedepth (\texttt{max\_treedepth}) that should be used. More on these control settings can be found at \url{https://mc-stan.org/misc/warnings.html}.  

\begin{spverbatim}
    m_O <- brm(O | trunc(lb=10, ub = 50) ~ female + (1|age_group), 
    data=data_us, chains=4, cores=4, control=list(adapt_delta=.99))
\end{spverbatim}

\noindent
In math this can be written as

\begin{gather}
    y_i \sim normal(\beta_0 + \beta_{\operatorname{male}}X_{\operatorname{female[i]}}+\alpha_{age[i]},\sigma) \\
    \alpha_{age} \sim normal(0,\sigma_{age})
\end{gather}
One important feature of the Bayesian workflow is the selection of priors. By default brms normalizes and rescales the data and sets priors that reflect this transformation.  We can change the default prior choices using the {\tt prior} argument to the {\tt brm} call.

One way of understanding the choice of priors is using prior predictive checks \parencite{gabry2019visualization}. Using an additional argument we can sample from the prior only. One thing to note is that the default prior in brms for a $\beta$ parameter is unconstrained, which is difficult to sample from. We specify a wide $N(0,10)$ prior to enable us to do a prior predictive check, but to not advocate for this prior necessarily in all situations. 

\begin{spverbatim}
    m_O <- brm(O | trunc(lb=10, ub = 50) ~ female + (1|age_group), 
    data=data_us, chains=4, cores=4, control=list(adapt_delta=.99), 
    prior = set_prior("normal(0,10)", class = "b"), sample_prior = "only")

\end{spverbatim}

\begin{figure}
    \centering
    \includegraphics[width=12cm]{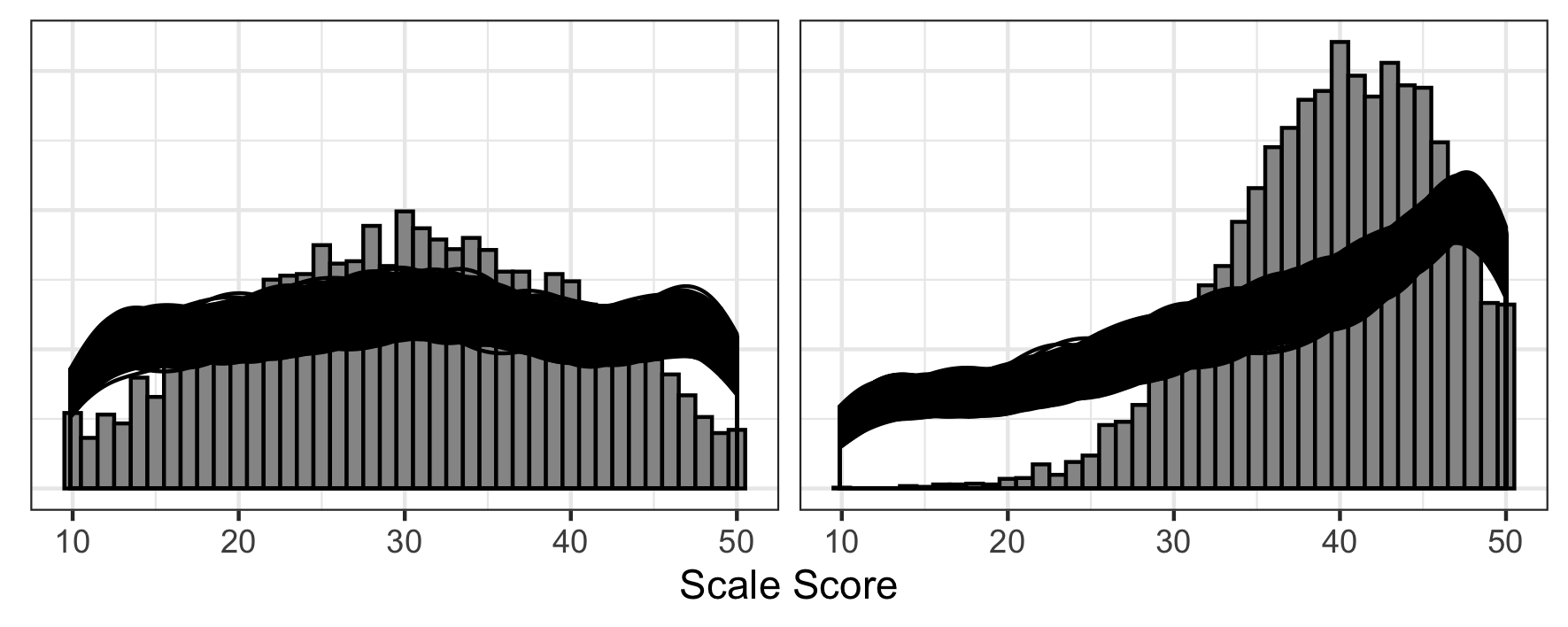}
    \caption{Observed sample histogram for each of the five personality subscales. For each we display the posterior estimates for the sample (black lines) and population (red), which give an indication of uncertainty of our estimates.}
    \label{fig:oe_pp}
\end{figure}

In Figure~\ref{fig:oe_pp} we plot the expected distribution for the Openness and Extraversion given the model. These priors are not updated by the data, but because they are created relative to normalized data, they have different effects given different outcomes. For a more thorough description of Bayesian workflow in psychological examples, refer to \textcite{schad2019principled}.

This model fits well with no warnings. The focus of this manuscript is not on how to test good fitting in Bayesian models so we do not discuss this further here.  We direct readers towards \textcite{gabry2019visualization} for more tools on effective model checking and diagnostics. The  takeaway from this step is that we have used our sample to fit an estimate of O scores for different gender and age groups. We plot the estimate for each in Table~\ref{tab:param_ests}. Other models could also have been used for this.

\begin{table}[ht]
\caption{Parameter estimates for the intercept if male, females, and six different age categories. }
\centering
\label{tab:param_ests}
\begin{tabular}{lrrrr}
\toprule
  & Posterior Mean & Posterior sd & Quantile 2.5 & Quantile 97.5\\
\midrule
intercept & 43.1 & 0.4 & 42.4 & 43.9\\
female & -2.7 & 0.2 & -3.1 & -2.3\\
\addlinespace
<18 & -0.6 & 0.4 & -1.5 & 0.1\\
18-24 & -0.5 & 0.4 & -1.3 & 0.2\\
25-34 & 0.4 & 0.4 & -0.3 & 1.2\\
35-45 & 0.3 & 0.4 & -0.5 & 1.1\\
45-64 & 0.4 & 0.4 & -0.4 & 1.2\\
65+ & 0.0 & 0.5 & -0.9 & 1.0\\
\bottomrule
\end{tabular}
\end{table}

\subsection{Step 2: Adjust the sample to the population}

Next we need an estimate for the population distribution of the adjustment variables, in this case age and gender. We would like to generalize to the population of U.S. residents aged 13 and over (the youngest participant in the survey is aged 13). We get the population distribution of age $\times$ gender from the American Community Survey \parencite[ACS]{ACS}, a large representative survey of the US, which we can use with the provided weights to approximate census level information. We use ACS estimates from 2012, the year when most of our sample data were collected. This is a large, supplementary survey to the population wide decennial census that is conducted yearly. It comes with a set of weights that are used to obtain population level estimates of quantities measured. 

After downloading and merging the files (the ACS is released in four datafiles), we subset down to the age and gender variables. Using the age variable, we create the same age categories as we used in the sample. We can then use the ACS survey weights to estimate the number of people in each combination of age group and gender. We use the package dplyr for this, and print the resulting poststratification matrix in Table~\ref{tab:acs_ps}.

\begin{table}
\centering
\caption{Population counts of each combination of demographics as estimated using the ACS, where $N$ is the number of Americans in that category.}
\label{tab:acs_ps}
    \begin{tabular}{ccc}
    \toprule
    female & age\_group & N\\
    \midrule
    0 & 1 & 10\,713\,479\\
    0 & 2 & 15\,974\,402\\
    0 & 3 & 22\,216\,888\\
    0 & 4 & 20\,279\,699\\
    0 & 5 & 31\,659\,960\\
    0 & 6 & 30\,275\,386\\
    \addlinespace
    1 & 1 & 10\,193\,764\\
    1 & 2 & 15\,166\,108\\
    1 & 3 & 21\,758\,629\\
    1 & 4 & 20\,455\,441\\
    1 & 5 & 32\,907\,478\\
    1 & 6 & 36\,732\,643\\
    \bottomrule
    \end{tabular}
\end{table}

After fitting this model for the openness score, we simulate a random sample of size $10\,000$ from the population, proportional to the estimated population cell sizes. Any sample size could be used, depending on the desired precision.

\begin{spverbatim}
    sample_pop <- sample(1:12, 10000,prob=acs_ps$N, replace=TRUE)
    sample <- acs_ps[sample_pop,1:2]    
\end{spverbatim}

\noindent
We then use a function from the brms package to predict the Opennness scores we would have observed in this simulated sample. We use the following code to estimate five possible Openness scores for each person in the sample. More could be taken, but we do this get a sense of posterior variance. 

\begin{spverbatim}
    PPC_O <- posterior_predict(m_O, newdata = sample)
\end{spverbatim}

\begin{figure}
    \centering
    \includegraphics[width=12cm]{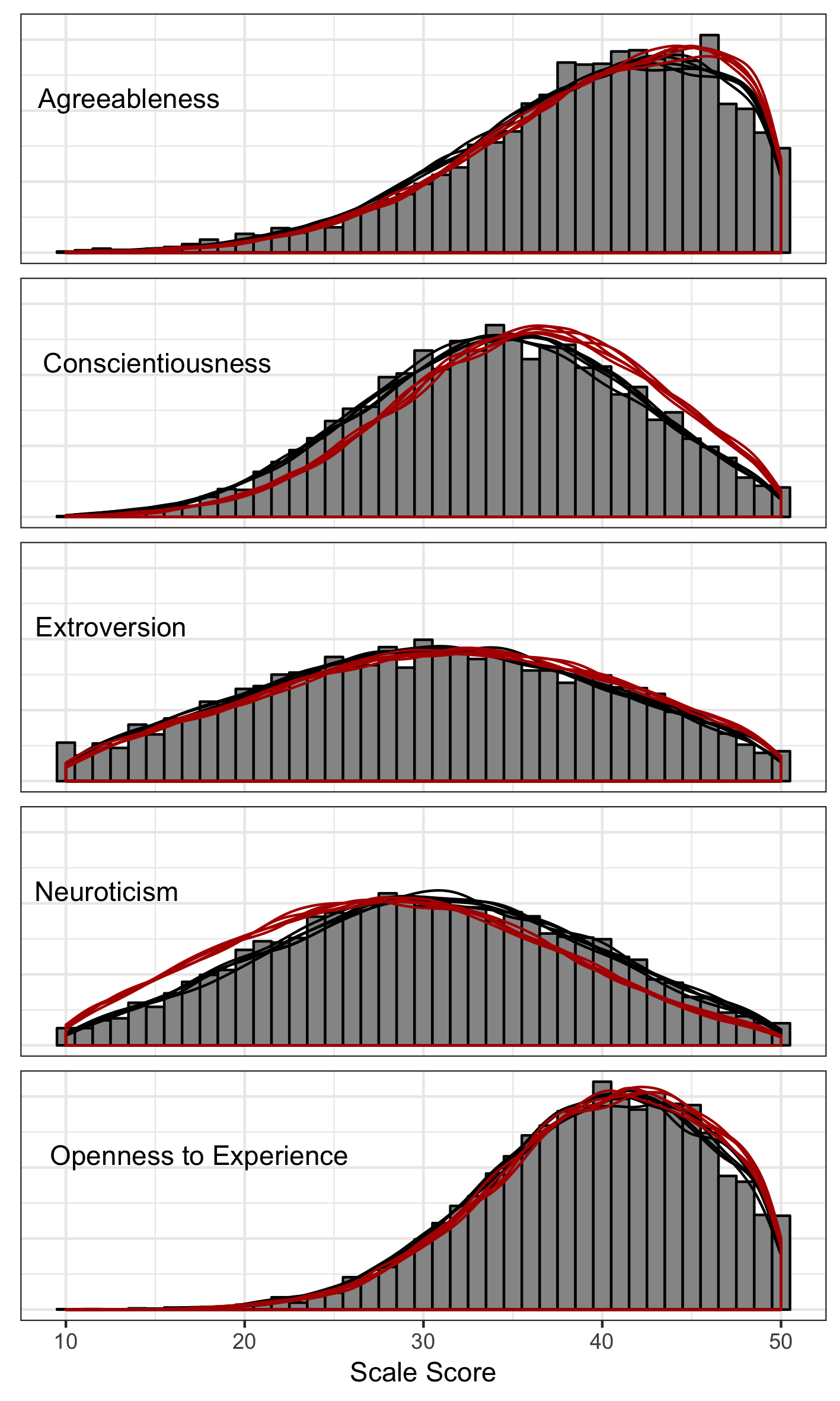}
    \caption{Observed sample histogram for each of the five personality subscales. For each we display the posterior estimates for the sample (black lines) and population (red), which give an indication of uncertainty of our estimates.}
    \label{fig:ocean_est}
\end{figure}

\noindent
We use a similar line of code to predict for the observed data to compare the predicted distributions of the model given the sample. In Figure~\ref{fig:ocean_est} we plot the sample (histogram), sample estimates (black lines) and population estimates (red lines). There are multiple lines to represent each posterior predictive estimate, giving an indication of uncertainty of our estimates. We can see that this MRP adjustment makes a considerable adjustment for some subscales (such as conscientiousness and neuroticism), a small amount of difference for others (openness and agreeableness) and negligible difference for extraversion. 

\subsection{Constraints on this analysis}

Using multilevel regression and poststratification in this analysis we used a non-probability convenience sample to estimate the population distributions of different psychometric scales. In our analysis we adjusted for age and gender, noting that the sample differed considerably from the population on these two demographics. This analysis is limited in that we did not adjust for education (as it was not measured), and we are concerned that education could be related to various personality factors and (judging from the dominant age/gender of the sample) we suspect that many of the respondents where psychology undergraduates. We also did not adjust for race/ethnicity (which is a common adjustment variable in political science) because of substantial differences in measurement between the sample and the US census. Lastly we focused only on individuals who responded as either male or female due to data constraints in the census. The purpose of our analysis is to provide an open data tutorial and proof of conscience rather than using these curves as a gold standard going forward.

\section{Example 2: Simulated experimental data}

Here we present a fictional but plausible example to extend this idea to experimental psychology. Say you would like to estimate impact of an intervention on maths anxiety.

For convenience, you ask students from your first year psychology class to take a survey that measures maths anxiety as well as a selection of demographic such as like age, gender, and major field of study. You also post flyers inviting participants from other faculties to participate, but your sample is not representative of the distribution of degree or gender at the university. Following the initial survey, participants are allocated to an intervention designed to reduce maths anxiety, or a control task. After they have completed, they are asked complete the maths anxiety survey again.

Writing a COG statement you acknowledge that while all students were members of the population, the sample was \textit{not} representative of the population of interest (the body of undergraduate students at your university). Furthermore, given that there might be interactions of gender or major with maths anxiety (maths majors might be less likely to be maths anxious than psychology majors), you declare that the total maths anxiety estimate from your sample might not be representative of the undergraduate population as a whole. In addition, you declare that gender or major might interact with the efficacy of the intervention, and so the estimate of the effect of the intervention might not be representative of the intervention's effects of the undergraduate population. 

There are multiple possible aims for generalizing this study. One aim might be to estimate the degree of maths anxiety that exists in the university. Another might be to estimate the impact of the maths anxiety intervention if it were implemented across the university. A third aim might be to replicate the study's results with a new sample from the same university. We address each of these aims in turn to explore some of the potential for MRP in psychology. 

\subsection{Estimating maths anxiety in the university}

The COG provides the framework to identify key areas where the sample differs from the population and how this might impact the results. What it doesn't do is provide a way of estimating maths anxiety in the actual population of interest. MRP partnered with the COG statement, however, provides a way to estimate maths anxiety in the full undergraduate population from the sample, without additional data collection. The procedure to do so would be as follows:

\begin{enumerate}
    \item Measure gender and degree major in the initial survey. In the simulated data that accompanies this tutorial we assume the initial sample is a generous n=300. 
    \item Obtain demographic data about the full population of undergraduate students at your university. This may or may not be easy, but we assume that undergraduate demographic data are published by or available from your university. Use the demographic data from the population to construct a poststratification table. This  table counts how many people in each possible demographic category (i.e., the number of women studying for an engineering degree, the number of men studying for an economics degree, etc.). The table should look something like the following, with the N column summing to the total number in the undergraduate population. It should contain all possible combinations of the demographic categories, but some may be empty. 
    \begin{center}
    \begin{tabular}{c c c } 
     \hline
     Gender & Major & N\\
     \hline
     F & Engineering & 982 \\ 
     F & Law & 1392 \\ 
     \vdots & \vdots & \vdots \\
     M & Liberal Arts & 672 \\
     M & Liberal Arts & 342 \\
     \vdots & \vdots & \vdots\\
     \hline
    \end{tabular}
    \end{center}
    \item Identify the outcome variable that you're interested in; here, it is baseline maths anxiety at a university level. 
    \item Using the sample, create a multilevel model with the demographic variables (especially gender and major) as predictors and and maths anxiety as the outcome variable. In our case we provide simulated data assuming the maths anxiety scale ranges from 10 (low) to 50 (high) so that we can use the same priors in brms as before. The only slight difference is that because this is simulated data we can include move beyond a binary gender variable if the data show a need for such an analysis. We fit the model using

\begin{spverbatim}
    brm(mathsanxiety_t1 | trunc(lb=10, ub = 50) ~ (1|gender) + 
    (1|major), data=dat_s1, chains=4, cores=4, 
    control=list(adapt_delta=.99),
    prior = set_prior("normal(0,10)", class = "b"))
\end{spverbatim}

    \item Use the model from 4 to predict the degree of maths anxiety using the poststratification table from 2. As with the first example, we do this for both the sample and the population, taking numerous posterior draws to compare the noise of the estimates. Unlike the previous example, we have much less data and so the estimates of distribution are much noisier. Also in contrast to the first example, our simulated population is much smaller as well, only $4222$ in this university. This means we can predict pre-intervention maths anxiety for each individual in the population. We plot the sample and population in Figure~\ref{fig:premath}.
    
    \begin{figure}
    \centering
    \includegraphics[width=12cm]{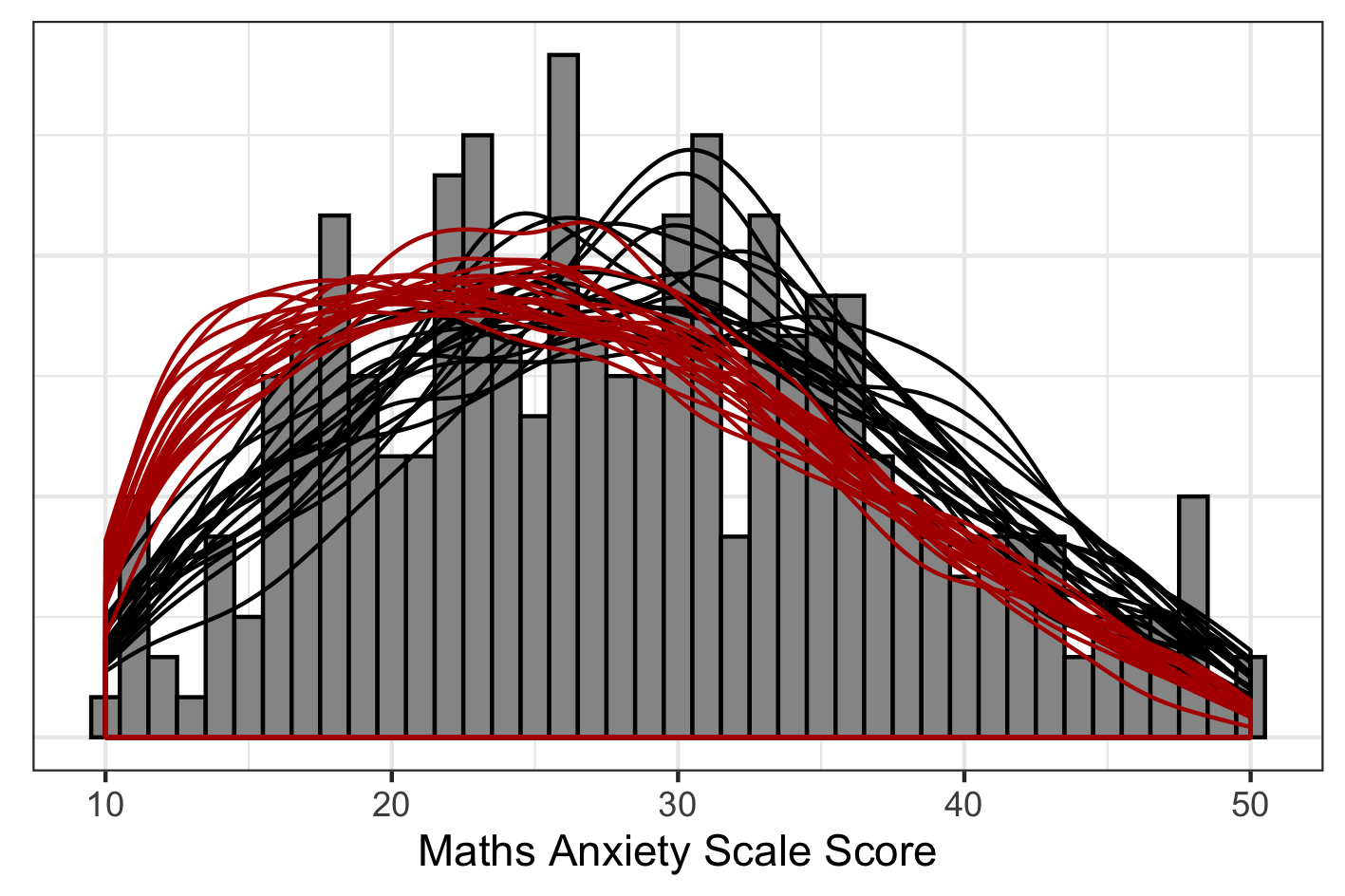}
    \caption{Observed sample histogram for the simulated sample maths anxiety scale score. For each we display multiple posterior estimates for the sample (black lines) and population (red), which give an indication of uncertainty of our estimates.}
    \label{fig:premath}
\end{figure}

    \item Aggregate over these estimates using the size of the cell to estimate population or subpopulation values. In this case we calculate the average maths anxiety in the sample as M = 28.4 we estimate math anxiety in the university as M = 25.9. As this is a simulated dataset, we know the true mean of maths anxiety in that university is 26.7. 
\end{enumerate}

\subsection{Estimating the impact of the maths anxiety intervention in the university}

Often the primary aim of psychological research is not simply to estimate a quantity but instead to estimate the impact of an intervention or manipulation. To do so we often rely on random assignment to intervention and control groups. However, if the sample is different to the population, even randomization doesn't guarantee that the effect estimated in the sample will generalize to the wider population. Here we extend our MRP analysis to include randomized control trials. Building off the analysis presented in the previous section, we start at step 3.

\begin{enumerate}
  \setcounter{enumi}{2}
  
  \item We now use the sample to predict the post-intervention (Z where Z=0 if control group and Z=1 if treated) maths anxiety given the pre-intervention anxiety level, gender and major of the participant. This means that we are simply changing item 3 in the previous item to be the difference between pre and post maths anxiety scores.

    \item Now using the sample we model the difference between pre and post intervention maths anxiety for the control and intervention groups. If we were able to model maths anxiety using a linear model, then we would be able to model the before-afer difference directly as the difference between two normal distributions is normally distributed. As we are using a truncated regression to model maths anxiety, we instead model preintervention anxiety given gender and major
  
  \begin{spverbatim}
    brm(mathsanxiety_t1 | trunc(lb=10, ub = 50) ~ (1|gender) +
    (1|major), data=dat_s1, chains=4, cores=4, 
    control=list(adapt_delta=.99),
    prior = set_prior("normal(0,10)", class = "b"))
\end{spverbatim}
    and then post treatment anxiety given gender and major and pre treatment anxiety
    
\begin{spverbatim}
    brm(mathsanxiety_t2 | trunc(lb=10, ub = 50) ~ mathsanxiety_t1 + 
    (Z|gender) + (Z|major), data=dat_s1, 
    chains=4, cores=4, control=list(adapt_delta=.99),
    prior = set_prior("normal(0,10)", class = "b"))
\end{spverbatim}

    \item Then we use the first model to predict the degree of pre-intervention maths anxiety for each undergraduate in the university (taking 20 posterior samples to maintain uncertainty) and then for each posterior predicted estimate for math anxiety before treatment we can predict the post intervention predicted estimate for math anxiety as if each individual was allocated to either treatment or control.
    
    \item Using these two estimates, we can calculate the expected difference between pre and post maths anxiety given treatment and control intervention for both sample and population. We plot the estimated difference in both sample and undergraduate population in Figure~\ref{fig:mathdiff}. The mean post-pre difference in the sample is -3.36 for the intervention group and -0.47 for the control group. We estimate it in the undergraduate population as -4.66  for the intervention group and 0.11 for the control condition. As this is simulated data we also know the true effect in the population is -4.04 in the intervention group and 0.16 in the control. 
    
\begin{figure}
    \centering
    \includegraphics[width=12cm]{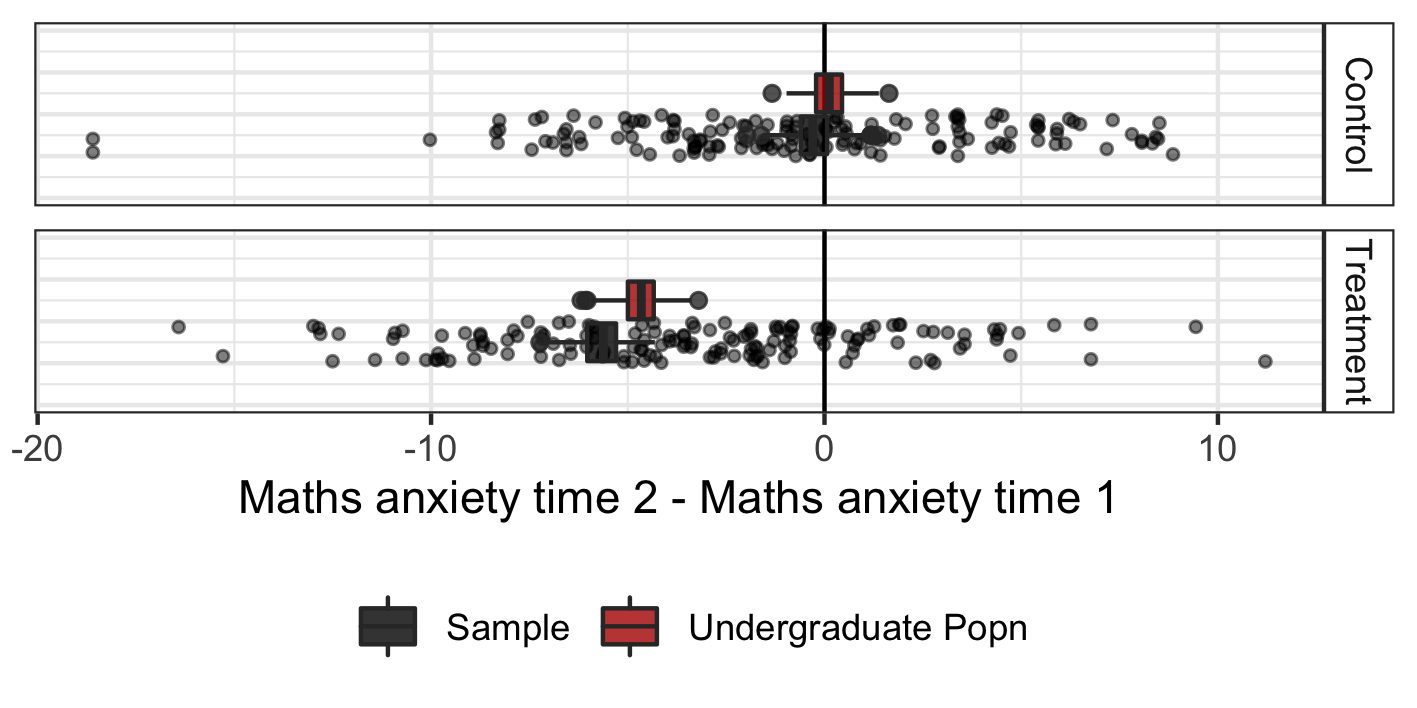}
    \caption{Expected post-pre difference in maths anxiety estimated in both the population (red) and the sample (black). Points represent observed differences in the sample.}
    \label{fig:mathdiff}
\end{figure}

\end{enumerate}

\subsection{Generalize the impact of the maths anxiety intervention in a new sample in the university}

Using the COG statement to explicitly define the population in terms of several key demographic features provided us the opportunity to make estimates for the population. However, \textcite{simons2017constraints} noted that the purpose of the COG statement was more than simply describing the population. It also provides an avenue for future researchers to estimate the degree to which they ought to replicate the findings with a new sample based on the features of the current sample.

Say you are interested in the difference between pre and post treatment for an intervention. In your sample, you find a mean difference of $c$. Another researcher attempts to replicate your intervention with the same population, but finds their estimate of the difference to be $d$, where $d$ is of the opposite sign to $c$. However, the two samples differ on a number of demographic variables. The question is whether you ought to expect to see a difference $d$ in the sample given that you saw a difference $c$ in the original sample. 

Following on from our previous example of a maths anxiety intervention, we now consider using our first sample, which happens to contain mostly social science students, to predict the expected impact of the intervention on another sample now mostly engineering and science students. If we expect there to be heterogeneity in the expected effect of the intervention, it is possible that the effect observed in sample 1 will be different than the raw effect observed in sample 2. 

\begin{enumerate}
  \setcounter{enumi}{1}
    \item In this case we are going to treat sample 2 as the population. We could summarize it as a poststratification table, but provided it is relatively small (in this case we simulate a second sample of 300) we can simply use it as individual data.
    \addtocounter{enumi}{2}
    We can repeat steps three and 4 from the previous section using sample 1. 
    \item Now we can predict the baseline maths anxiety and post maths anxiety in sample 2 pretending as though each participant was given both treatment and control. This is similar to step 5 in the previous section, except the population is now the second sample. 
    \item We can repeat Figure~\ref{fig:mathdiff} but instead of predicting the difference between treatments in the undergraduate population, we predict the difference in treatment and control in sample 2 using sample 1. We present these estimates in Figure~\ref{fig:mathdiff2}. 
    
    \begin{figure}
    \centering
    \includegraphics[width=12cm]{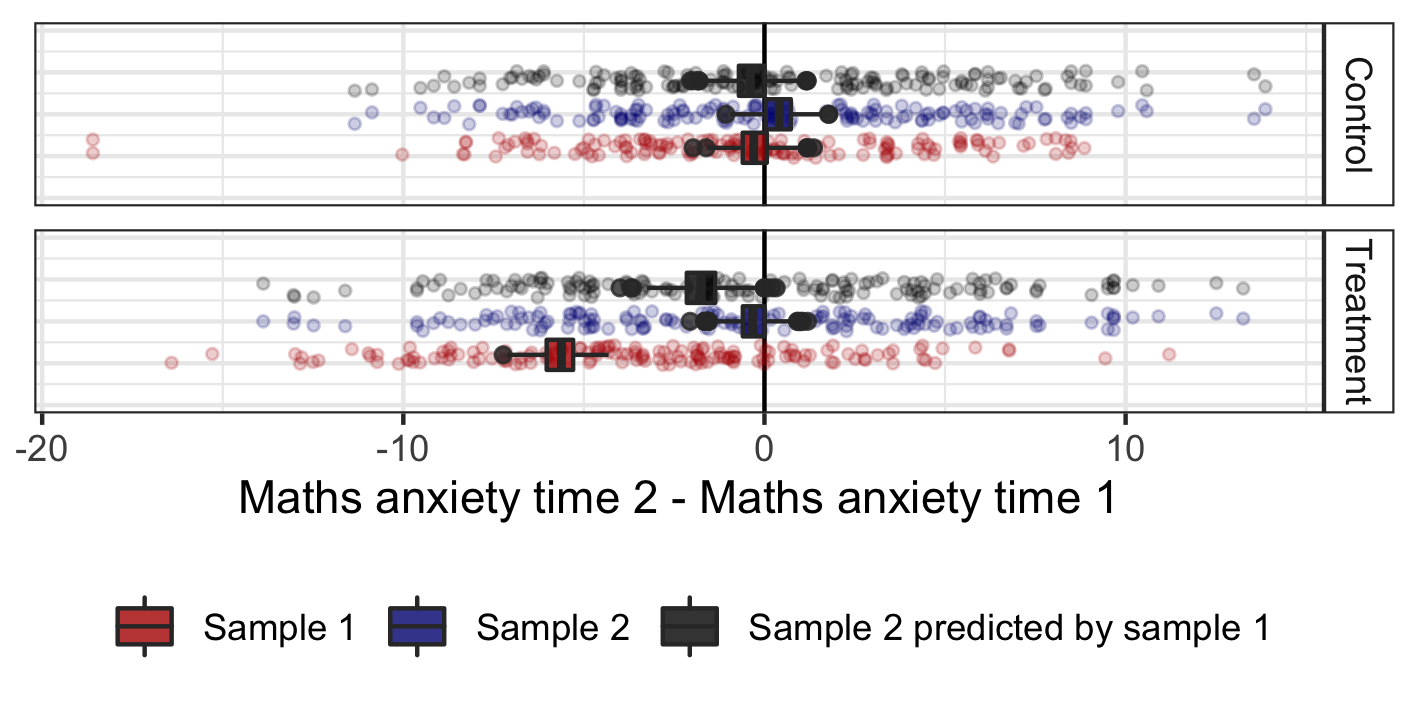}
    \caption{Expected post-pre difference in maths anxiety estimated in sample 1 (red), sample 2 (blue) and sample 1 predicting sample 2 (black). Points represent observed differences in the sample (sample 2 in the first two rows in each figure, sample 1 in the third).}
    \label{fig:mathdiff2}
\end{figure}
\end{enumerate}

\subsection{Generalize to other universities}

In the data accompanying this tutorial, we simulate not just one university but multiple universities, subsetting down to one for simplicity's sake in the previous sections. If we wanted to generalize to other universities then unless we were to assume that the effect across universities were the same after adjusting for other demographics, then we would need to have samples from multiple universities and model university, and university characteristics as another random effect in the model. An example of a sampling design that aimed to account for heterogeneity between schools, see \textcite{yeager2019national}. For simplicity we do not go through this here; the approach would likely be similar to approaches by \textcite{lax2009gay} to model state. 

\section{Active research areas}

At this point you may notice that all of these examples of MRP's possibilities share certain features.  Absence of these features correspond to some of the limitations of our method. MRP is widely used throughout the political science literature but is relatively new to psychology. The provided examples demonstrate that it already can be a valuable tool, but research need to be done to modify this method to specifically suit psychology's aims. 

One of the main challenges of MRP as set up above is that it is designed to estimate a parameter in the population given some demographic characteristics. In a pre-post design, the difference between pre and post can be treated as the outcome and implemented similarly or completed in a two-stage process as demonstrated in this tutorial. However, psychology is a science that considers complex relationships. For instance, consider the example used by \textcite{simons2017constraints} for the article by \textcite{whitsett2014approach} investigating the relationship between support seekers distress and willingness to provide support, mediated by high and low personal distress. In their COG statement, \textcite{simons2017constraints} note that the sample was ``a large number of different undergraduates sampled from the subject pool at the University of Washington'' and that they ``believe the results will be reproducible with students from similar subject pools serving as participants.''

From this information, we can infer that new undergraduates sampled from the University of Washington would be expected to show a similar relationship between participant willingness to support and support seeker's degree of distress. But further modeling would be needed to formalize this in an MRP context, which would allow the relationship to change given demographic characteristics. We expect the approach would be similar to \textcite{hill2011bayesian}.

Indeed, all the examples in the paper by \textcite{simons2017constraints} involve the generalization of an observed relationship in a sample to a wider population or a different sample. The Dunning-Kroger example concerns a replication to player groups with a more diverse skill level, while the suggested memory effect COG involves potential replications to investigate the effect of changes to experimental design. 

We believe there is a place for MRP to answer these types of questions, but we cannot point to an article that demonstrates this approach, and we suspect that further research needs to be done on the choice of sensible priors to induce regularization in a reasonable way. 

\section{Limitations}

As always in statistics, our claims are only as good as our models. MRP (or, more generally, regularized regression and poststratification) relies on a model or procedure to predict the outcome variable given some set of demographics. This model can fail to make good predictions for several reasons, including insufficient data, the lack of some demographic predictor, misspecification of some important part of the model, or insufficient regularization. Partial pooling in multilevel modeling uses data efficiently to mitigate some of these concerns, and a solid COG statement helps to provide some focus on the others.

Other limitations to this method arise because not only do we need to collect demographic variables in the sample (arguably relatively easy to do with some forethought), but we also need estimates of these same demographic variables in the population. These data are often available through government and census data, but not always and not always in the desired form. Some creativity may be needed to coerce available data into the desired form. For example, in political surveys it can be helpful to poststratify on party identification, which is not in the census and so one must use other surveys to estimate its distribution conditional on the relevant demographic and geographic predictors.

Moving forward, we would like to adjust and average not just for demographics but also for situation.  After all, effects in psychology experiments typically vary by situation, indeed that's often what is being studied. In this case there may be no clearly defined population distribution (what would it mean to define a population probability for being focused on a task, for example?), and so it would make sense to define some sort of artificial distribution representing the generalization of interest.  The requirement to specify such a distribution, like the COG, is itself salutory in forcing the researcher to think about the ultimate inferential goals:  not just learning from the data, but stating when this would apply.

\section{Conclusion}

We argue here that one of the important benefits of the COG statement is that it paves the way for statistical methods like MRP to be used, and we encourage other researchers to join us in considering how and when this technique might benefit the field.

Psychology has developed methods of estimating and evaluating internal validity through our rich and rigorous training in experimental design. However, we must not let a stellar job of accounting for internal validity distract us from also considering external validity. Inferences from convenience and snowball samples have serious threats to external validity once we consider heterogeneity of effects.

We have demonstrated how psychology can use MRP to estimate average treatment effects in defined populations, a particularly relevant task when working with non-probability samples. However, MRP will not always be a perfect solution. MRP is useful for adjusting a non-representative sample to a larger population. It is not, however, designed for situations where there are no individuals in a particular sub-population present in the sample (for example, using data from a WEIRD sample to generalize to the larger population of the world). In this case, we must either rely on strong assumptions or broaden our data pool through collaborations across the world---which is perhaps one of the most encouraging possibilities for MRP.  

\printbibliography{}

\end{document}